\documentclass{article}
\usepackage{authblk}
\usepackage{amsmath,amssymb,amsfonts,amsthm}
\usepackage{array}
\usepackage[caption=false,font=normalsize,labelfont=sf,textfont=sf]{subfig}
\usepackage{textcomp}
\usepackage{stfloats}
\usepackage{url}
\usepackage{verbatim}
\usepackage{graphicx}
\usepackage{cite}
\usepackage[T1]{fontenc}
\usepackage{thmtools}
\usepackage{thm-restate}
\usepackage{algorithmicx}
\usepackage{algorithm}
\usepackage{algpseudocode}
\usepackage{graphicx}
\usepackage{booktabs}
\usepackage{tikz}
\usepackage{standalone}
\usepackage{accents}
\newtheorem{theorem}{Theorem}

\newtheorem{remark}{Remark}

\newcommand\shl{\mathrel{\scriptstyle{\mathtt{{<}{<}}}}}
\newcommand\shr{\mathrel{\scriptstyle\tt\mathtt{{>}{>}}}}

\begin{document}

\title{Efficiency of ANS Entropy Encoders}

%\author{\textsc{Dmitry Kosolobov}\\
%	\normalsize Ural Federal University, Ekaterinburg, Russia\\
%	\normalsize dkosolobov@mail.ru
%}
%\date{}

\author[1,2]{Dmitry Kosolobov}% <-this % stops a space
%	\thanks{
\affil[1]{St.~Petersburg State University, St.~Petersburg, Russia.}
\affil[2]{Ural Federal University, Ekaterinburg, Russia.   E-mail:~dkosolobov@mail.ru. ORCID: 0000-0002-2909-2952}

\date{}


%\markboth{IEEE Transactions on Information Theory}{Efficiency of ANS Entropy Encoders}

%\IEEEpubid{0000--0000~\copyright~2023 IEEE}


\maketitle

\begin{abstract}%\boldmath
Asymmetric Numeral Systems (ANS) is a class of entropy encoders that had an immense impact on the data compression, substituting arithmetic and Huffman coding. It was studied by different authors but the precise asymptotics of its redundancy (in relation to the entropy) was not completely understood. We obtain asymptotically tight bounds for the redundancy of the tabled ANS (tANS), the most popular ANS variant. Given a sequence $a_1,a_2,\ldots,a_n$ of symbols from an alphabet $\{0,1,\ldots,\sigma-1\}$ such that each symbol $a$ occurs in it $f_a$ times and $n=2^r$, the tANS encoder using Duda's ``precise initialization'' to fill tANS tables transforms this sequence into a bit string of the following length (the frequencies are not included in the encoding):
$\sum\limits_{a\in[0..\sigma)}f_a\cdot\log\frac{n}{f_a}+O(\sigma+r)$,
where $O(\sigma+r)$ can be bounded by $\sigma\log e+r$. The $r$-bit term is an artifact indispensable to ANS; the rest incurs a redundancy of $O(\frac{\sigma}{n})$ bits per symbol. We complement this by examples showing that an $\Omega(\sigma+r)$ redundancy is necessary. We argue that similar examples exist for most adequate initialization methods for tANS. Thus, we refute Duda's conjecture that the redundancy is $O(\frac{\sigma}{n^2})$ bits per symbol.
We also propose a variant of the range ANS (rANS), called rANS with fixed accuracy, parameterized by $k\ge 1$ that in certain conditions might be faster than the standard rANS because it avoids slow explicit division operations.
We bound the redundancy for our rANS variant by $\frac{n}{2^k-1}\log e+r+k$.

\noindent\textbf{Keywords:} 	Asymmetric numeral systems, ANS, finite state entropy, FSE, arithmetic coding, redundancy
\end{abstract}

%\begin{IEEEkeywords}
%	Asymmetric numeral systems, ANS, finite state entropy, FSE, arithmetic coding, redundancy.
%\end{IEEEkeywords}

\algtext*{EndFunction}% Remove "end function" text
\algtext*{EndProcedure}% Remove "end procedure" text
\algtext*{EndFor}% Remove "end for" text
\algtext*{EndWhile}% Remove "end while" text
\algtext*{EndIf}% Remove "end if" text
\algloopdefx{NIf}[1]{\textbf{if} #1 \textbf{then}}
\algloopdefx{NElse}{\textbf{else}}
\algloopdefx{NElseIf}{\textbf{else if}}
\algloopdefx{NForAll}[1]{\textbf{for each} #1 \textbf{do}}
\algloopdefx{NWhile}[1]{\textbf{while} #1 \textbf{do}}
\algloopdefx{NFor}[1]{\textbf{for} #1 \textbf{do}}

\section{Introduction}

\emph{Asymmetric numeral systems (ANS)} is a class of entropy encoders invented by Duda in 2009~\cite{Duda2009,Duda2013,DudaEtAl}. These encoders had a huge impact on the data compression by providing the same rates of compression as the arithmetic coding~\cite{Martin,MoffatNealWitten,Pasco,Rissanen,WittenNealCleary} while being as fast as the Huffman coding~\cite{Huffman} (and even faster in some scenarios). Since the invention of ANS and the emergence of its efficient implementation by Collet~\cite{Collet}, several high performance compressors based on ANS appeared~\cite{LZFSE,ColletZstd,Giesen} and it was integrated in some modern media formats \cite{AlakuijalaEtAl,AlonsoSutterDeVergata}. The theoretical community also contributed to the study of ANS in a series of works~\cite{BlanesEtAL,Duda2021,MoffatPetri,WangWangLin,YamamotoIwata,Yokoo,DubeYokoo,YamamotoIwata2}, though less actively than practitioners~\cite{Bloom,ColletBlog,GiesenBlog,Giesen}.

The primary focus of the theoretical analysis for entropy encoders is in the estimation of the redundancy, the difference between the number of bits produced by the encoder and the information theoretic entropy lower bound. The main result of the present paper is a tight asymptotic upper and lower bound on the redundancy of the most popular variant of ANS, called the tabled ANS (tANS) or, sometimes, the Finite State Entropy (FSE). Our analysis of the redundancy does not impose any randomized assumptions on the source that produced the input for the encoder; instead, we establish worst-case bounds in terms of the so-called \emph{empirical entropy}~\cite{KosarajuManzini,Manzini} (the definitions follow). In addition, we also introduce and analyze a novel variant of the range ANS (rANS), another version of ANS used in practice. 

An entropy encoder receives as its input a sequence of $n$ numbers from a set $\{0,1,\ldots,\sigma-1\}$ and transforms it into a bit string. Duda analyzed his tANS encoder and found that its redundancy is $O(\frac{\sigma}{n}\log n)$ bits per symbol\footnote{All logarithms in the paper are in base two.} provided an appropriate initialization is used for tANS tables (details are explained in the sequel); see Equation~(40) in~\cite{Duda2009}. His estimation, however, seems somewhat non-rigorous and in a different (more classical) setting: he considers a memoryless source generating the input at random and he estimates the expected length of the output encoding. Based on experimental evaluations, Duda conjectured that a tight bound for the redundancy is $O(\frac{\sigma}{n^2})$ bits per symbol~\cite{Duda2013}. Yokoo and Dub{\'e}~\cite{DubeYokoo} investigated the same problem in more rigorous terms closer to our setting and showed that the redundancy per symbol vanishes as the length $n$ tends to infinity while $\sigma$ is fixed (yet, they have some questionable assumptions in their derivations). Another analysis of the expected output length for the memoryless source was conducted in~\cite{YamamotoIwata}: their bounds, however, are incomplete in a sense and, thus, cannot be easily compared to other results. In contrast, our analysis establishes upper bounds in the worst case without probabilistic assumptions.

In this paper we prove that the redundancy for tANS is $O(\frac{\sigma}{n})$ bits per symbol. We complement this upper bound by a series of examples showing that it is asymptotically tight when $\sigma > n/3$. As in the works cited above, here we did not include in this bound the $r = \log n$ redundant bits that are always produced by ANS encoders in the worst case (it is an artifact of the initial state of the encoder). After uncovering the constant under the big-O and including this $r$-bit term, the upper bound for the tANS redundancy that we establish can be expressed as $\sigma\log e + r$ bits over all symbols (not per symbol). Our lower bound examples show that a redundancy of $\frac{\sigma - 1}{4} + r - 2$ bits is attainable, for $\sigma \approx n/3$.
An important part of tANS is the initialization of its internal tables, which has a significant impact on the compression performance. The lower bound examples work for a wide class of initialization algorithms so that the same redundancy can be observed for any adequate algorithm that generates the tANS tables using only the known frequencies of symbols without processing the sequence itself (however, methods that analyze the sequence might avoid this effect~\cite{DubeYokooDist,Duda2021}; these methods are usually infeasible in practice due to their relatively slow performance). We note that an analogous upper bound of $O(\frac{1}{2^w})$ bits per symbol is known for practical variants of the arithmetic coding that use $w$-bit machine words \cite{HowardVitter}.

For the tANS encoder, it is natural to assume that the number of its internal states, $n = 2^r$, is comparable with the length of the input sequence since the tANS uses quite large internal tables for good performance. All above results are stated for the case when the length of the input sequence, $m$, is equal to $n$. When it is not the case and $m$ is larger than $n$, we also obtain upper bounds; in particular, we rederive a very recent result from \cite{YamamotoIwata2}: when $n$ and $\sigma$ are fixed and $m$ tends to infinity, the redundancy per symbol can be upperbounded by $D(p||\hat{p}) + O(\frac{\sigma}{n})$, where $D(p||\hat{p})$ is the Kullback--Leibler divergence between the empirical distribution of symbols, $p$, and the approximate distribution, $\hat{p}$, that is used by the tANS with $n$ internal states (details follow).

The second contribution of the present paper is a modification of the range ANS (rANS). The rANS is another variant of ANS invented by Duda. It has a number of advantages in some use scenarios because of which it was favoured by some practitioners (and because it is easier to learn). The rANS was first noticeably slower in practice than the tANS but its current SIMD implementations tend to be at least on a par with tANS. The main advantage of rANS is that it does not need tANS tables (however, fast variants still require some specific tables). Due to the less wasteful use of memory, the rANS is more suitable for (pseudo)adaptive compression or when several streams of data are encoded simultaneously~\cite{Bloom,GiesenBlog,Giesen,Strutz}. An inherent problem of rANS that slows down it significantly is a necessity to perform the integer division during the encoding, which is expensive on modern processors. The proposed modification of rANS, which we call rANS with fixed accuracy, tries to mitigate this issue making the encoder faster while preserving the same good properties of rANS.

The new rANS takes as a regulated parameter an integer $k \ge 1$. It is guaranteed that the division can be performed only in cases when its result belongs to the range $[2^k..2^{k+1})$. Thus, the division can be computed by faster methods provided $k$ is small. We upperbound the redundancy for the rANS with fixed accuracy $k$ by $\frac{n}{2^k - 1}\log e + r$. In our experiments, the new rANS variant with $k=3$ provides faster encoding than the standard rANS with explicit divisions but slower than the rANS in which the divisions are performed by multiplications and shifts on precomputed constants (see \cite{Alverson,LemireBartlettKaser}); the decoding in the new rANS was slower. We believe that the rANS with fixed accuracy might be suitable when the encoder should be adaptive (so that one cannot precompute constants for faster divisions) and the encoding speed is a priority.

The paper is organized as follows. After short preliminaries, Section~\ref{sec:ans} introduces a simple variant of the range ANS and, based on it, the tabled ANS. Section~\ref{sec:analysis} presents a tight analysis of the redundancy for the tabled ANS, upper and lower bounds. Section~\ref{sec:enhanced-rans} describes a novel variant of range ANS and analyzes it. We conclude with some open problems in Section~\ref{sec:conclusion}.

\section{Preliminaries}

A \emph{symbol} is an element of a finite set of integers called an \emph{alphabet}. We consider sequences of symbols. The sequences are also sometimes called \emph{strings}. Integer intervals are denoted by $[i..j] = \{k \in \mathbb{Z} \colon i\le k\le j\}$ and $[i..j) = [i..j]\setminus\{j\}$. Fix an alphabet $[0..\sigma)$ with $\sigma$ symbols and a sequence of symbols $a_1, a_2, \ldots, a_n$ from it. For each symbol $a$, denote by $f_a$ the number of occurrences of $a$ in the sequence. The number $f_a$ is called the \emph{frequency of $a$}. The number $f_a / n$ is the \emph{empirical probability of $a$}. An entropy encoder transforms this sequence into a sequence of bits that is then transmitted to a decoder. The encoder usually also transmits to the decoder the table of frequencies $f_a$. However, the problem of storage for the table is out of the scope of the present paper and we focus only on the encoding for the sequence itself implicitly assuming that both sides know the length $n$ of the sequence and the frequencies of symbols.
The following quantity is used as an information theoretic lower bound~\cite{CoverThomas,KosarajuManzini} for the number of bits that any encoder should spend to encode some (least compressible) sequences with the given frequencies of symbols $\{f_a\}_{a\in [0..\sigma)}$ (even under the assumption that the table of frequencies $f_a$ is known):
\begin{equation}
\sum\limits_{a \in [0..\sigma)} f_a \log\frac{n}{f_a},\label{eq:entropy}
\end{equation}
assuming that $f_a \log\frac{n}{f_a} = 0$ whenever $f_a = 0$. 
We call~\eqref{eq:entropy} the \emph{entropy formula} (though the empirically calculated entropy itself, i.e., the \emph{empirical entropy}, is defined as the quantity~\eqref{eq:entropy} divided by $n$). The difference between the number of bits produced by an encoder and the optimal number of bits from~\eqref{eq:entropy} is called a \emph{redundancy}. The redundancy is the primary focus of our analysis of ANS.

For each symbol $a$, denote $c_a = \sum_{a' \in [0..a)} f_{a'}$ (these numbers are called \emph{cumulative frequencies}). Typically, encoders store two arrays of $\sigma$ numbers $c_a$ and $f_a$ and some additional tables necessary for encoding. The ANS encoders are not an exception.

\begin{remark}
We digress at this point to discuss how the quantity~\eqref{eq:entropy} can be interpreted as a lower bound. As it was mentioned in the introduction, we aim to avoid any probabilistic assumptions on the source of the input. In this case, the information theoretic lower bound for the length of the encoding is \begin{equation}
	\left\lceil\log\binom{n}{f_0,f_1,\ldots,f_{\sigma-1}}\right\rceil = \left\lceil\log\frac{n!}{f_0! f_1!\cdots f_{\sigma-1}!}\right\rceil, \label{eq:multinom}
\end{equation}
provided the length $n$ of the sequence and the table of frequencies $f_a$ are accessible to the encoder and decoder for free. The bound \eqref{eq:multinom} is tight: it is achieved by a simple encoder that uses the same number of bits as in \eqref{eq:multinom} for all inputs whose symbol frequencies are $f_0,f_1,\ldots,f_{\sigma-1}$, and the encoding is the index of the input in the lexicographically sorted list of all possible inputs. The tight bound \eqref{eq:multinom} is less than the entropy bound~\eqref{eq:entropy} but close (e.g., see \cite[Lem. II.2]{Csiszar}): the difference is at most $O(\sigma\log\frac{n}{\sigma})$.  Hence, \eqref{eq:entropy} is not a lower bound in this strict sense.

The quantity \eqref{eq:entropy} is equal to $\min\log(1/\Pr(Q\text{ produces }a_1,a_2,\ldots,a_n))$, where the minimum is taken for all memoryless sources $Q$ randomly generating sequences of length $n$ over the alphabet $[0..\sigma)$ (see~\cite{Gagie}). If one restricts attention only to encoders that produce uniquely decodable codes (note that the encoder produces one codeword for the whole sequence of length $n$ from the alphabet $[0..\sigma)$, not symbol by symbol), then \eqref{eq:entropy} is a lower bound for the expected number of bits produced by such encoder provided each symbol of the length-$n$ input sequence is generated by the memoryless source with respective probabilities $f_a / n$ of the symbols~\cite{McMillan,CoverThomas}. It is this fundamental fact that seems to justify the usage of~\eqref{eq:entropy} as a lower bound in~\cite{KosarajuManzini,Manzini} and numerous subsequent papers. (The lower bound for non-uniquely decodable codes is close to~\eqref{eq:entropy} but smaller; see \cite{AlonOrlitsky,SzpankowskiVerdu}.)

It is still convenient to refer to~\eqref{eq:entropy} as a kind of ``optimal lower bound'' and, following the established tradition, we call it as such in the sequel, despite the fact that it is not exacty the case. %\cite{OrlitskySanthanamZhang} 

As a side note: if the table of frequencies cannot be accessed by the decoder for free and, thus, should be transmitted somehow, then the tight lower bound for the length of the uniquely decodable code is equal to \eqref{eq:entropy} plus $\Theta(\sigma\log\frac{n}{\sigma})$ when $\sigma < n$ (see~\cite{OrlitskySanthanam,Shtarkov,SzpankowskiWeinberger}). Therefore, \eqref{eq:entropy} is indeed a lower bound in this case but it is not tight (recall that this is not our setting: we consider the frequencies as freely transmitted to the decoder).
\end{remark}

\section{ANS Encoders}\label{sec:ans}

On a high level, an ANS encoder can be described as a data structure that maintains a positive integer $w$ that can be modified by two stack operations ``$w = \mathsf{push}(w, a)$'' and ``$(w,a) = \mathsf{pop}(w)$'': $\mathsf{push}$ encodes a symbol $a$ into the number $w$ and returns the modified value for $w$, and $\mathsf{pop}$ performs a reverse operation decoding the symbol $a$ and restoring the old value for $w$. It is instructive to image the number $w$ as a sequence of bits (potentially very long) from the highest bit of $w$, which is always~1, to lowest bits. Given a sequence of symbols $a_1, a_2, \ldots, a_n$, the encoder pushes them consecutively and transmits the resulting integer $w$ to a decoder, which in turn retrieves the sequence from $w$ performing pop operations. Note that the symbols appear in the reverse order during the decoding, which is a distinctive feature of ANS.
For their correct coordinated work, both encoder and decoder should receive in advance the same table of (approximate) frequencies of symbols in the sequence $a_1, a_2, \ldots, a_n$.

For didactic reasons, ANS encoders are usually first described in their unbounded form in which they operate on very long integers $w$, which is unrealistic. In practice, \emph{streaming ANS encoders} are used, which maintain the value of $w$ during the construction within a fixed range fitting into a machine word and they store all excessive bits in an output buffer by performing a ``renormalization'' (akin to arithmetic encoders). However, in what follows, we significantly diverge from this standard way of explanation for ANS and introduce only the streaming ANS but in an unconventional manner; the unbounded ANS variant is not discussed at all. The unconventional description of ANS was chosen since it fits more naturally in our upper bound analysis of ANS.

\subsection{Preliminary ANS}\label{sec:rans}

Suppose that we are to encode a sequence of symbols $a_1, a_2, \ldots, a_n$ from the alphabet $[0..\sigma)$. For each symbol $a \in [0..\sigma)$, denote by $f_a$ its frequency in the sequence (i.e., its empirical probability is $\frac{f_a}{n}$). We assume that $n$ is a power of two, i.e., $n = 2^r$ for an integer $r \ge 0$. It is a standard assumption for both ANS and arithmetic coding that simplifies implementations. (If the length of the sequence is not a power of two, then the real empirical probabilities of symbols are approximated with numbers of the form $f_a / 2^r$, where $2^r$ is a power of two close to $n$, $f_a$ are not necessarily real frequencies of symbols, and $\sum_a f_a = 2^r$; we discuss this case briefly in the end of Section~\ref{sec:analysis-upper-bound} but most details are omitted as they are not in the scope of this paper.)

The streaming ANS encoder reads the symbols $a_1, a_2, \ldots, a_n$ from left to right and, after processing $a_1, a_2, \ldots, a_i$, encodes the processed part into a positive integer $w_i$. Initially, $w_0 = 2^r$; the choice for $w_0$ is somewhat arbitrary, the only necessary condition is $w_0 \ge 2^r$. To encode a new symbol $a_{i+1}$, we transform the number $w_i$ into $w_{i+1}$ by replacing at most $r + 1$ highest bits of $w_i$ with exactly $r + 1$ bits that store a larger value than what was stored in the replaced bits; see Figure~\ref{fig:replace}. Thus, $w_0 < w_1 < \cdots < w_{i+1}$. According to our terminology, we have $w_{i+1} = \mathsf{push}(w_i, a_{i+1})$. We view the number $w_i$ as a bit stream, from highest to lowest bits of $w$.
\begin{figure}[hb]
$$
\begin{array}{rr}
w_i =&\underline{10010}\,011001000010110\\
w_{i+1}=&\underbrace{1001011}_{r + 1\text{ bits}}011001000010110
\end{array}
$$
\caption{A transformation of the number $w_i$ into $w_{i+1}$.}\label{fig:replace}
\end{figure}

The numbers $w_i$ might be very long. However, since only the highest $r + 1$ bits of $w_i$ matter for the encoder, all lower bits can be stored in an output buffer. The integer $r$ is chosen so that the $r + 1$ bits can be stored into one machine word. For simplicity of the exposition, we omit this technical detail and continue to discuss the numbers $w_i$ as if they were stored explicitly.

The goal is to encode the whole sequence optimally so that, ideally, the final number $w_n$ occupies $\sum_a f_a \log\frac{n}{f_a}$ bits. Intuitively, one can achieve this by encoding each symbol $a$ into $\log\frac{n}{f_a} = r - \log f_a$ bits. During the processing of $a = a_{i+1}$, we could have achieved this by replacing the highest $\log f_a + 1$ bits of $w_i$ with $r + 1$ bits that store a new larger value; then, the number of bits in $w_{i+1}$ and $w_i$ will differ by $r + 1 - (\log f_a + 1) = \log\frac{n}{f_a}$. But the number $\log f_a$ is not integer in general. Therefore, instead, we replace either $\lfloor\log f_a\rfloor + 1$ or $\lfloor\log f_a\rfloor + 2$ highest bits of $w_i$ with new $r + 1$ bits. As a result, the number of bits in $w_{i+1}$ increases by either $r - \lfloor\log f_a\rfloor$ or $r - \lfloor\log f_a\rfloor - 1$, which is approximately $\log\frac{n}{f_a}$. The cumulative growth of $w_n$ may approach the optimal $\log\frac{n}{f_a}$ bits per symbol $a$ on average if the case when $\lfloor\log f_a\rfloor + 2$ bits are replaced happens more often. 
%(Note that when $f_a \ge n/2 = 2^{r-1}$, we have $\lfloor\log f_a\rfloor = r - 1$ and the number of bits in $w_i$ sometimes might not change at all; but the content will change.)

What is the content of the $r + 1$ new highest bits in $w_{i+1}$ and how do we decide whether $\lfloor\log f_a\rfloor + 1$ or $\lfloor\log f_a\rfloor + 2$ highest bits of $w_i$ will be replaced?

Denote by $x'$ the value stored in the highest $r+1$ bits of $w_{i+1}$, i.e., $w_{i+1} = x'\cdot 2^{\lfloor \log w_{i+1}\rfloor - r} + \Delta$, where $0 \le \Delta < 2^{\lfloor \log w_{i+1}\rfloor - r}$ and $2^r \le x' < 2^{r+1}$. Denote by $x$ the value of the highest bits of $w_i$ that were replaced with $x'$, i.e., $w_{i} = x\cdot 2^{\lfloor \log w_{i+1}\rfloor - r} + \Delta$, where $x \ge 1$. (In the example of Figure~\ref{fig:replace} $x$ and $x'$ are emphasized and $\Delta$ is a common part of $w_i$ and $w_{i+1}$.)
The scheme must be reversible: the number $x'$ must provide sufficient information for the decoder to restore the symbol $a_{i+1}$ and the number $x$. Since $\sigma \le \sum_a f_a = 2^r$, there is enough place to encode $a_{i+1}$ into the $r$ lowest bits of $x'$ (the highest, $(r+1)$th, bit of $x'$ is~$1$). ``Excessive'' $2^r - \sigma$ possible values of $x'$ will be used to restore the number $x$, the replaced highest bits of $w_i$.

The $r$ lowest bits of $x'$ can store any number from the range $[0..2^r)$. The encoder chooses one of these numbers depending on the values of $x$ and $a_{i+1}$. We distribute the numbers from $[0..2^r)$ among the symbols according to their frequencies: for each symbol $a$, denote $c_a = \sum_{a' \in [0..a)} f_{a'}$ (a cumulative frequency corresponding to $a$); the subrange of values $[c_a..c_{a+1})$ is allocated for $a$. Hence, given a symbol $a = a_{i+1}$, we have a room in the range that can store any number from $[0..f_a)$ and this information should suffice to restore the replaced number $x$ from $x'$. As an example, Figure~\ref{fig:distribution} depicts a distribution of the range $2^r = 2^4$ among symbols $a, b, c, d$ with frequencies $3, 5, 6, 2$, respectively.

\begin{figure}[H]
$$
\begin{array}{ll}
\begin{array}{|c|c|c|c|c|c|c|c|c|c|c|c|c|c|c|c|}
\hline
a & a & a & b & b & b & b & b & c & c & c & c & c & c & d & d \\
\hline
\end{array} & ~\\
0 & \!\!\!\!\!2^r
\end{array}
$$
\caption{A distribution of symbols in a range of length $2^r$.}\label{fig:distribution}
\end{figure}

Let $a = a_{i+1}$. Note that the number $f_a$ occupies $\lfloor\log f_a\rfloor + 1$ bits. Denote by $x_1$ and $x_2$ the values stored in, respectively, the highest $\lfloor\log f_a\rfloor + 1$ and the highest $\lfloor\log f_a\rfloor + 2$ bits of $w_i$. We assume that $x = x_1$ if $x_1 \ge f_a$, and $x = x_2$ otherwise (see Figure~\ref{fig:cases}, where the bits occupied by $x_1$ and $x_2$ are emphasized, respectively, on the left and right). Thus, the condition $x_1 \ge f_a$ determines whether $\lfloor\log f_a\rfloor + 1$ or $\lfloor\log f_a\rfloor + 2$  highest bits of $w_i$ will be used for $x$. Note that $2^{\lfloor\log f_a\rfloor} \le x_1 < 2\cdot f_a$ and $f_a < 2^{\lfloor\log f_a\rfloor+1} \le x_2 \le 2\cdot x_1 + 1$. If $x = x_1$ (i.e., $f_a \le x_1$), we have $f_a \le x < 2\cdot f_a$. If $x = x_2$ (i.e., $f_a > x_1$), we have $x_2 \le 2\cdot x_1 + 1 < 2\cdot f_a$. Therefore, both cases imply $f_a \le x < 2\cdot f_a$. Thus, $x \bmod f_a = x - f_a$.
\begin{figure*}[htb]
$$
\begin{array}{rlcrl}
w_i =&\underline{10110}011001000010110 & ~~~ & w_i =&\underline{100100}11001000010110\\
f_a =&10011                            & ~~~ & f_a =&\phantom{1}10011\\
x \bmod f_a =&\phantom{111}11          & ~~~ & x \bmod f_a =&\phantom{1}10001
\end{array}
$$
\caption{Two cases: $x$ occupies $\lfloor\log f_a\rfloor + 1$ highest bits of $w_i$ (left) or $\lfloor\log f_a\rfloor + 2$ highest bits of $w_i$ (right).}\label{fig:cases}
\end{figure*}

In order to restore the value $x$ from $x'$, it suffices to encode somehow the value $x \bmod f_a$ into $x'$: then, $x$ is restored as $x = f_a + (x \bmod f_a)$. To this end, the range $[c_a..c_{a+1})$ allocated for the symbol $a$ has exactly enough room. Thus, we have the following transformation to encode $a$:
\begin{equation}
x' = 2^r + c_a + (x \bmod f_a),\text{ where }f_a \le x < 2\cdot f_a.\label{eq:encode}
\end{equation}
In other words, the operation $w_{i+1} = \mathsf{push}(w_i,a)$ assigns to a variable $x$ the highest $\lfloor\log f_a\rfloor + 1$ or $\lfloor\log f_a\rfloor + 2$ bits of $w_i$ (depending on the content of the bits, as was described above) and replaces these bits by $x'$ computed in \eqref{eq:encode}, where $x \bmod f_a = x - f_a$, thus transforming $w_{i}$ into $w_{i+1}$. (In practice, instead of the entire $w_i$, the encoder stores only the highest $r + 1$ bits of $w_i$, which suffices for these calculations.)

With this approach, the decoder should perform a reverse transformation for the $r + 1$ highest bits of the number $w_{i+1}$ in order to restore $w_i$. The decoding is straightforward:
\begin{equation}
x = f_a + (x' \bmod 2^r) - c_a,\text{ where }c_a \le x' \bmod 2^r < c_{a+1}.\label{eq:decode}
\end{equation}
The decoded symbol $a$ is determined by examining to which range $[c_a..c_{a+1})$ the number $x' \bmod 2^r$ belongs. This is how the operation ``$(w_i,a_{i+1}) = \mathsf{pop}(w_{i+1})$'' is performed. It remains to observe that $x' > x$ since $x' \ge 2^r + (x \bmod f_a) > f_a + (x \bmod f_a) = x$. Therefore, the number $w_{i+1}$ indeed is larger than $w_i$.

\iffalse
To summarize, let us give a pseudocode for the described procedures.
\begin{algorithmic}[1]
\Function{$\mathsf{push}$}{$x, a$} \Comment{$x$ occupies $r + 1$ bits}
    \State $k = k[a] - [x < \mathsf{bound}[a]];$ \Comment{$[c]$ is $1$ if $c$ is true, $0$ otherwise}\label{lst:condition}
    \State $\mathsf{outputBits}(x, k)$;
    \State\Return{$2^r + c_a + ((x \shr k) - f_a);$}
\EndFunction
\end{algorithmic}
\begin{algorithmic}[1]
\Function{$\mathsf{pop}$}{$x'$} \Comment{$x$ occupies $r + 1$ bits}
    \State $a = \mathsf{symbol}[x' \bmod 2^r];$
    \State $x = f_a + (x' \bmod 2^r) - c_a;$
    \State $\mathsf{readBits}(x, k)$;
    \State\Return{$(x,a);$}
\EndFunction
\end{algorithmic}
Line~\ref{lst:condition} in $\mathsf{push}$ was simplified by Collet using a neat trick: $k = (x + t[a]) \shr r$, where $t[a]$ stores the number .
\fi

\medskip

The described scheme is the simplest preliminary form of the so-called \emph{range ANS (rANS)}. As will be seen later, the size of $w_n$ in bits can be bounded by $\sum_a f_a \log\frac{n}{f_a} + O(n)$. The redundancy $O(n)$ is quite significant for many applications and, indeed, the described preliminary scheme does not perform well in practice and it is never used as is. To fix this issue, an additional ``shuffling'' step should be added to the encoder, which is described in the following section; the scheme with this step is called \emph{tabled ANS (tANS)} or \emph{Finite State Entropy (FSE)} and it has much smaller redundancy. There exists, however, a more elaborate version of the range ANS that also has a small redundancy and works well without shuffling; we discuss it briefly in Section~\ref{sec:enhanced-rans} where we also present a novel variant of the range ANS with good compression guarantees.

\subsection{Shuffling and Tabled ANS}

The idea of the shuffling step enhancing the described preliminary ANS scheme is to shuffle the lower bits of $x'$ in a random-like fashion. Thereby, the scheme~\eqref{eq:encode} is changed as follows:
\begin{equation}
x' = 2^r + \mathsf{shuffle}[c_a + (x \bmod f_a)],\text{ where }f_a \le x < 2\cdot f_a.\label{eq:encode-shuffle}
\end{equation}
The array $\mathsf{shuffle}[0..2^r{-}1]$ is a permutation of the range $[0..2^r)$. Despite its name, the permutation is initialized not randomly but by a fixed deterministic algorithm, a \emph{shuffling method}. Some methods are described in the next subsection. Each of the methods gives rise to a different ANS variant. Some of them use a pseudo-random generator with a fixed seed shared by the encoder and decoder, which somewhat justifies the name ``shuffling''. The permutation $\mathsf{shuffle}$ cannot be entirely random: in order to guarantee the inequality $x' > x$ (which implies $w_{i+1} > w_i$), it must satisfy the following property:
\begin{equation}
\mathsf{shuffle}[c_a + j] < \mathsf{shuffle}[c_a + k],\text{ whenever }0 \le j < k < f_a.\label{eq:decode-shuffle}
\end{equation}
Due to this condition, we have $x' > x$ since $x' \ge 2^r + (x \bmod f_a) > f_a + (x \bmod f_a) = x$. It is convenient to view $\mathsf{shuffle}$ as defined via an additional array $\mathsf{range}[0..2^r{-}1]$ that stores an (arbitrary) permutation of the array of symbols from Figure~\ref{fig:distribution} in which every symbol $a$ occurs exactly $f_a$ times; then, for $j \in [0..f_a)$, $\mathsf{shuffle}[c_a + j]$ is equal to the index of the $(j+1)$th occurrence of the symbol $a$ in $\mathsf{range}$ (see Figure~\ref{fig:random-dist}). Thus, to define $\mathsf{shuffle}$, it suffices to initialize the array $\mathsf{range}$; we will implicitly imply this relation in the sequel when the initialization of $\mathsf{shuffle}$ is discussed.

\newcommand{\rn}[2]{%% "rn": "remember node"
    \tikz[remember picture,baseline=(#1.base)]\node [inner sep=0] (#1) {$#2$};%
}

\begin{figure}[H]
$$
\begingroup
\begin{array}{rll}
~ &
\begin{array}{|c|c|c|c|c|c|c|c|c|c|c|c|c|c|c|c|}
\hline
a & a & a & \rn{b1}{b} & \rn{b2}{b} & \rn{b3}{b} & \rn{b4}{b} & \rn{b5}{b} & c & c & c & c & c & c & d & d \\
\hline
\end{array} & ~\\
~ & 0 & \!\!\!\!\!2^r \\
\mathsf{range} = &
\begin{array}{|c|c|c|c|c|c|c|c|c|c|c|c|c|c|c|c|}
\hline
\rn{bb1}{b} & d & a & c & c & \rn{bb2}{b} & a & \rn{bb3}{b} & c & d & \rn{bb4}{b} & c & c & \rn{bb5}{b} & c & a \\
\hline
\end{array} & ~\\
~ & 0 & \!\!\!\!\!2^r
\end{array}
\endgroup
\begin{tikzpicture}[overlay,remember picture]
\draw [->] (b1) -- (bb1);\draw [->] (b2) -- (bb2);\draw [->] (b3) -- (bb3);\draw [->] (b4) -- (bb4);\draw [->] (b5) -- (bb5);
\end{tikzpicture}
$$
\caption{A shuffled distribution of symbols. Here, we have $\mathsf{shuffle}[c_b] = 0, \mathsf{shuffle}[c_b+1] = 5, \mathsf{shuffle}[c_b+2] = 7$.}\label{fig:random-dist}
\end{figure}

The decoding procedure~\eqref{eq:decode} performs a reverse transformation in an obvious way:
$$
x = f_a + \mathsf{unshuffle}[x' \bmod 2^r] - c_a,\text{ where }\mathsf{range}[x' \bmod 2^r] = a.
$$
Note the use of the array $\mathsf{range}$ in the decoder to determine the symbol $a$.
The array $\mathsf{unshuffle}$ is the inverse of $\mathsf{shuffle}$ such that $\mathsf{unshuffle}[\mathsf{shuffle}[z]] = z$, for any $z \in [0..2^r)$. Thus, it ``moves'' the value $x' \bmod 2^r$ to its ``correct'' location and we have to add the frequency $f_a$ and subtract the cumulative frequency $c_a$ afterward. However, implementations usually construct, instead of the arrays $\mathsf{unshuffle}$ and $\mathsf{range}$, an array $\mathsf{decode}$ that stores, for each number $x' \bmod 2^r$, an already corrected value for $x$ and the corresponding symbol $a$; hence, the decoding is much simpler:
$$
(x,a) = \mathsf{decode}[x' \bmod 2^r].
$$

The described scheme is called a \emph{tabled ANS (tANS)}. It is the most popular variant of ANS widely used in practice. The choice of the shuffling method is crucial for its performance. Some methods are considered in the next section. There are several additional technical improvements that can be applied to this basic scheme. Perhaps, the most notable of them is that one can feed to the decoder more than $r + 1$ bits at once, decoding many symbols in one step (the information about the decoded symbols and the new value for $x$ must be stored in the array $\mathsf{decode}$). Also, we point out again that $x \bmod f_a$ in~\eqref{eq:encode-shuffle} is computed as $x - f_a$ since $f_a \le x < 2\cdot f_a$. We do not discuss these details further.

\subsection{Shuffling Methods}\label{sec:shuffling-methods}

The general rule for shuffling is to distribute symbols in the array $\mathsf{range}$ as uniformly as possible so that, for any symbol $a$, the distance between consecutive occurrences of $a$ is approximately $\frac{n}{f_a}$. Implementations usually use heuristics for this~\cite{Collet} or Duda's method~\cite{Duda2013} (which is introduced below).
Let us discuss some considerations on this regard that will be developed in a more rigorous way in Section~\ref{sec:analysis}.

The following informal argument suggests why the scheme with shuffling enhances the ANS so that it achieves the entropy bound (the rigorous analysis is in Section~\ref{sec:analysis}). The value $\log x' - \log x$ is, in a sense, an increase in bits from the number $w_i$ to $w_{i+1}$; the bit length of $w_n$, the final encoding, is approximately the sum of the increases (this reasoning is formalized in Section~\ref{sec:analysis}). Denote $\delta = x \bmod f_a$. We have $x = f_a + \delta$. If the symbols in the array $\mathsf{range}$ are distributed uniformly, the distance between two consecutive symbols $a$ is approximately $\frac{n}{f_a}$. Therefore, using~\eqref{eq:encode-shuffle}, the encoder transforms $x$ approximately to $x' \approx 2^r + \frac{n}{f_a}\delta = \frac{n}{f_a}(f_a + \delta) = \frac{n}{f_a}x$ (recall that $n = 2^r$). Hence, we obtain $\log x' - \log x \approx \log\frac{n}{f_a}$, which is precisely the optimal number of bits for the symbol $a$ according to the entropy formula.

The argument suggests that the shuffling method should spread the symbols in the array $\mathsf{range}$ in such a way that the distance between two consecutive occurrences of $a$ is approximately $\frac{n}{f_a}$ and the encoder transforms the number $x = f_a + \delta$ as close as possible to the number $2^r + \frac{n}{f_a}\delta$. Under this assumption of ``uniformity'', if the first occurrence of symbol $a$ is at position $p$ in $\mathsf{range}$, then $x$ is transformed into $x' \approx 2^r + \frac{n}{f_a}\delta + p$. The term $p$ adds to the redundancy associated with $f_a$ symbols $a$: the larger the value of $p$, the more bits are spent per symbol $a$. Therefore, the first occurrences of more frequent symbols should be closer to the beginning of the array $\mathsf{range}$ so that they produce less redundancies overall. Duda's method, which he called a \emph{precise initialization}, tries to take into account all these considerations.

Duda's algorithm maintains a priority queue with the following operations: $\mathsf{put}(q, a)$ adds a pair of numbers $(q,a)$ to the queue; $(q,a) = \mathsf{getmin}()$ removes from the queue a pair $(q,a)$ with the smallest value $q$ (breaking ties arbitrarily). We first give in Algorithm~\ref{alg:init} simplified Duda's algorithm, which is easier to analyze.

\begin{algorithm}
\caption{A simple initialization algorithm.}\label{alg:init}
\begin{algorithmic}
\For{$a \in [0..\sigma)$}
    \State $\mathsf{put}(0, a),\, d_a = c_a;$
\EndFor
\For{$i = 0,1,\ldots,2^r-1$}
    \State $(q,a) = \mathsf{getmin}(), $
     $\mathsf{range}[i] = a, $
     $\mathsf{shuffle}[d_a] = i;$
    \State $\mathsf{put}(q + \frac{n}{f_a},a), $
     $d_a = d_a + 1;$
\EndFor
\end{algorithmic}
\end{algorithm}

The array $\mathsf{range}$ corresponding to $\mathsf{shuffle}$ is not needed for the encoder and its construction is added here for the convenience of the reader.

At every moment during the work of the algorithm, there is only one instance of each symbol in the priority queue and, if a symbol $a$ was assigned to the array $\mathsf{range}$ exactly $k$ times, then it is represented by the pair $(\frac{n}{f_a}k, a)$ in the queue. Therefore, after $f_a$ assignments of $a$ into $\mathsf{range}$, the symbol is represented by $(n,a)$ and all other symbols $b$ that had less than $f_b$ assignments in $\mathsf{range}$ are represented as $(q,b)$ with $q < n$. Hence, in the end, each symbol $a$ occurs in $\mathsf{range}$ exactly $f_a$ times.

Duda's original ``precise initialization'' is the same as Algorithm~\ref{alg:init} except that the operation ``$\mathsf{put}(0, a)$'' from the first loop is changed to ``$\mathsf{put}(\frac{1}{2}\cdot \frac{n}{f_a}, a)$''. Its correctness is proved analogously.

\section{Analysis of the Tabled ANS}\label{sec:analysis}

We are to estimate the redundancy of the output produced by the ANS encoder, i.e., the difference between $\lceil\log w_n\rceil$ and the lower bound $\sum_a f_a\log\frac{n}{f_a}$. We postpone the analysis of the ANS without shuffling to the next section, where its more general variant is considered. In this subsection we consider the ANS with shuffling, i.e., the tabled ANS (tANS). To the best of our knowledge, the following analysis, albeit quite simple, evaded the attention of researchers and was not present in prior works. Our proof methods, however, stem from observations and arguments from Dub{\'e} and Yokoo~\cite{DubeYokoo} and Duda~\cite{Duda2009,Duda2013}.

\subsection{Upper Bounds}\label{sec:analysis-upper-bound}

Let us upperbound $\log w_{i+1} - \log w_{i} = \log\frac{w_{i+1}}{w_{i}}$, a bit increase after one step of the encoding procedure. The total number of bits will be then estimated as follows (the term $r$ appears because $w_0 = 2^r$ and $\log w_0 = r$):
\begin{equation}
\log w_n = \log\frac{w_n}{w_{n-1}} + \log\frac{w_{n-1}}{w_{n-2}} + \cdots + \log\frac{w_1}{w_0} + r.\label{eq:wn}
\end{equation}

Suppose that $w_{i+1}$ was obtained from $w_{i}$ by ``inserting'' a symbol $a$ as described above. Denote $\ell = \lfloor \log w_{i+1}\rfloor - r$ so that $w_i = x \cdot 2^\ell + \Delta$ and $w_{i+1} = x' \cdot 2^{\ell} + \Delta$, where $0 \le \Delta < 2^\ell.$ Then, $\log\frac{w_{i+1}}{w_i} = \log\left(\frac{x' 2^\ell + \Delta}{x 2^\ell + \Delta}\right) = \log\left(\frac{x'}{x}\left(\frac{1 + \Delta / (x' 2^\ell)}{1 + \Delta / (x 2^\ell)}\right)\right) = \log x' - \log x + \log\left(\frac{1 + \Delta / (x' 2^\ell)}{1 + \Delta / (x 2^\ell)}\right)$. Since $x' > x$, the additive term $\log\left(\frac{1 + \Delta / (x' 2^\ell)}{1 + \Delta / (x 2^\ell)}\right)$ is non-positive and, thus,
we have obtained the following inequality:
\begin{equation}
\log w_{i+1} - \log w_i \le \log x' - \log x.\label{eq:w-to-x}
\end{equation}

It remains to estimate how close is $\log x' - \log x$ to the optimum $\log\frac{n}{f_a}$. We first consider the case when the encoder uses the shuffling produced by simplified Duda's algorithm (Algorithm~\ref{alg:init}).

Fix a symbol $a$ and a number $\delta \in [0..f_a)$. Denote by $k$ the index of the $(\delta+1)$th occurrence of $a$ in $\mathsf{range}$. Note that $\mathsf{shuffle}[c_a + \delta] = k$, by definition. For each $b \in [0..\sigma)$, denote by $k_b$ the number of symbols $b$ in the subrange $\mathsf{range}[0..k{-}1]$. Clearly, we have $k = \sum_{b} k_b$. The shuffling algorithm implies the following inequality:
\begin{equation}
(k_b - 1)\frac{n}{f_b} \le \delta \frac{n}{f_a}.\label{eq:kb-inequality}
\end{equation}
We express $k_b$ from~\eqref{eq:kb-inequality} as $k_b \le \delta\frac{f_b}{f_a} + 1$. Summing over all $b\in [0..\sigma)$, we deduce from this $k \le \delta\frac{n}{f_a} + \sigma$.
It follows from~\eqref{eq:w-to-x} that, in order to analyze the number of bits per symbol produced by the encoder, we have to estimate $\log x' - \log x$, where, by~\eqref{eq:encode-shuffle}, $x' = 2^r + \mathsf{shuffle}[c_a + (x \bmod f_a)]$. Assuming $\delta = x \bmod f_a$, we obtain $x = f_a + \delta$ and $x' = 2^r + k = n + k$. Therefore, $\log x' - \log x = \log(n + k) - \log x \le \log(n + \delta\frac{n}{f_a} + \sigma) - \log x = \log(\frac{n}{f_a} x + \sigma) - \log x = \log\frac{n}{f_a} + \log(1 + \frac{\sigma}{n x / f_a}) \le \log\frac{n}{f_a} + \frac{\sigma}{n}\log e$.
Thus, we estimate the number of bits per symbol $a$ as $\log\frac{n}{f_a} + \frac{\sigma}{n}\log e$, i.e., the redundancy is $\frac{\sigma}{n}\log e$ bits per symbols.

Now let us analyze Duda's original algorithm. The algorithm is the same as Algorithm~\ref{alg:init} except that the operation ``$\mathsf{put}(0, a)$'' from the first loop is changed to ``$\mathsf{put}(\frac{1}{2}\cdot \frac{n}{f_a}, a)$''. The analysis is slightly more complicated. First, an equation analogous to \eqref{eq:kb-inequality} for this case looks as follows:
$$
\left(k_b - \frac{1}{2}\right)\frac{n}{f_b} \le \left(\delta + \frac{1}{2}\right)\frac{n}{f_a}.
$$
We then similarly deduce $k_b \le (\delta + \frac{1}{2})\frac{f_b}{f_a} + \frac{1}{2}$ and, summing over all $b$, $k \le (\delta + \frac{1}{2})\frac{n}{f_a} + \frac{\sigma}{2}$. Again, assuming $x = f_a + \delta$, it follows from this that $\log x' - \log x \le \log(\frac{n}{f_a}x + \frac{n}{2f_a} + \frac{\sigma}{2}) - \log x = \log\frac{n}{f_a} + \log(1 + \frac{1}{2x} + \frac{\sigma}{2 nx / f_a}) \le \log\frac{n}{f_a} + (\frac{1}{2f_a} + \frac{\sigma}{2n})\log e$. The symbol $a$ occurs in the sequence exactly $f_a$ times. Hence, the redundancies $\frac{1}{2f_a}\log e$ for symbols $a$ sum up to $\frac{1}{2}\log e$ over all these occurrence and, therefore, in the end we obtain $\frac{\sigma}{2n}\log e$ bits per symbol contributed by the terms $\frac{1}{2f_a}\log e$, for all symbols $a$, in the final encoding. Adding to this the
$\frac{\sigma}{2n}\log e$ bits per symbol, we obtain $\frac{\sigma}{n}\log e$ redundant bits per symbol in the final encoding.

It remains to add the additive term $r$ from~\eqref{eq:wn} to the redundancy, which contributes $\frac{r}{n}$ bits per symbol, and the following theorem is proved.

\begin{theorem}
Given a sequence $a_1, a_2, \ldots, a_n$ of symbols from an alphabet $[0..\sigma)$ such that each symbol $a$ occurs in it $f_a$ times and $n = 2^r$ for an integer $r$, the ANS encoder using [simplified] Duda's precise initialization transforms this sequence into a bit string of length
$$
\sum\limits_{a \in [0..\sigma)} f_a \cdot \log\frac{n}{f_a} + O(\sigma + r),
$$
where the $O(\sigma + r)$ term can be bounded by $\sigma\log e + r = \sigma\log e + \log n$. Thus, we have $O(\frac{\sigma + r}{n}) = O(\frac{\sigma + \log n}{n})$ redundant bits per symbol.\label{thm:upper-bound}
\end{theorem}

In practice, the length $m$ of the encoded sequence $a_1, a_2, \ldots, a_m$ is not necessarily a power of two. A typical solution for this case is to approximate the real empirical probabilities $f_a / m$ of symbols with approximate ones $\hat{f}_a / 2^r$, where $\sum_a \hat{f}_a = 2^r$. The ANS encoder then processes the sequence as usually but using the ``frequencies'' $\hat{f}_a$ instead of $f_a$. The same analysis can be applied for this case: Equations~\eqref{eq:wn} and~\eqref{eq:w-to-x} trivially hold and the value $\log x' - \log x$ is bounded in the same manner by $\log\frac{2^r}{\hat{f}_a} + \frac{\sigma}{2^r}\log e$, for simplified Duda's initialization, and by $\log\frac{2^r}{\hat{f}_a} + (\frac{1}{2\hat{f}_a} + \frac{\sigma}{2\cdot 2^r})\log e$, for Duda's initialization. Summing the redundancies over all $m$ symbols, we obtain the following theorem (for simplicity, the theorem is stated only for simplified Duda's initialization).

\begin{theorem}
Let $a_1, a_2, \ldots, a_m$ be a sequence of symbols from an alphabet $[0..\sigma)$ such that each symbol $a$ occurs in it $f_a$ times. Let the probabilities $f_a / m$ be approximated by numbers $\hat{f}_a / n$ such that $\hat{f}_a$ are integers, $n = 2^r$, for an integer $r$, and $\sum_{a\in [0..\sigma)} \hat{f}_a = n$. The ANS encoder that uses the approximate probabilities and simplified Duda's initialization transforms this sequence into a bit string of length
$$
\sum\limits_{a \in [0..\sigma)} f_a \cdot \log\frac{n}{\hat{f}_a} + O(\frac{\sigma m}{n} + r),
$$
where the $O(\frac{\sigma m}{n} + r)$ term can be bounded by $\frac{\sigma m}{n}\log e + r$.\label{thm:upper-bound2}
\end{theorem}

Theorem~\ref{thm:upper-bound2} can be used as follows to estimate the redundancy per symbol in terms of the real empirical distribution $p$, given by the probabilities $p_a = \frac{f_a}{m}$, and the approximate distribution $\hat{p}$, given by $\hat{p}_a = \frac{\hat{f}_a}{n}$. Denote the entropy of $p$ and the Kullback--Leibler divergence between $p$ and $\hat{p}$ as:
 $$
 H(p) = \sum_{a\in [0..\sigma)} p_a \log\frac{1}{p_a},\quad\quad\quad D(p||\hat{p}) = \sum_{a\in [0..\sigma)} p_a \log\frac{p_a}{\hat{p}_a}
 $$
Then we derive:
\begin{flalign*}
& \sum\limits_{a \in [0..\sigma)} f_a \cdot \log\frac{n}{\hat{f}_a} + O(\frac{\sigma m}{n} + r) \\
&\quad\quad = m \left[ \sum\limits_{a \in [0..\sigma)} p_a \cdot \log\frac{1}{\hat{p}_a} + O(\frac{\sigma}{n} + \frac{r}{m}) \right]\\
&\quad\quad = m\left[ \sum\limits_{a \in [0..\sigma)} p_a \cdot \log\frac{1}{p_a} + \sum\limits_{a \in [0..\sigma)} p_a \cdot \log\frac{p_a}{\hat{p}_a} + O(\frac{\sigma}{n} + \frac{r}{m}) \right]\\
&\quad\quad = m\left[ H(p) + D(p||\hat{p}) + O(\frac{\sigma}{n} + \frac{r}{m}) \right].
\end{flalign*}
Thus, once $n$ and $\sigma$ are fixed and $m$ tends to infinity, the redundancy per symbol can be bounded by $D(p||\hat{p}) + O(\frac{\sigma}{n})$. This is the same estimation that was obtained by Yamamoto and Iwata in \cite{YamamotoIwata2} using different methods.

\subsection{Lower Bound Example}

Apparently, the $r$-bit redundancy incurred by the initial value $w_0$ is unavoidable in the described scheme. It is less clear whether an $O(\sigma)$ additive term is necessary in Theorem~\ref{thm:upper-bound}. An informal argument supporting that this is the case is as follows. Consider a sequence in which all symbols are (approximately) equiprobable, i.e., their frequencies $f_a$ are ${\sim}\frac{n}{\sigma}$. The lower bound for the encoding of the sequence is $n \log\sigma$ bits. The array $\mathsf{range}$ constructed by Duda's initialization algorithm for the sequence looks (approximately) as $n / \sigma$ blocks, each of which is of size $\sigma$ and consists of consecutive symbols $0, 1, \ldots, \sigma-1$. Hence, when the encoder receives a symbol $a_{i+1} = a \in [0..\sigma)$ during its work, it transforms the number $x = f_a + \delta$, where $\delta = x \bmod f_a$, occupying leading bits of $w_i$, into the number $x' = 2^r + \frac{n}{f_a}\delta + a = \frac{n}{f_a}x + a = \sigma x + a$ (note that $f_a = n / \sigma$, by our assumption). As in the previous section, one can deduce from this that $\log x' - \log x = \log\sigma + \log(1 + \frac{a}{\sigma x})$. Since $x < 2f_a = 2\frac{n}{\sigma}$ and $\frac{a}{2n} < 1$, the redundant additive term $\log(1 + \frac{a}{\sigma x})$ can be estimated as $\log(1 + \frac{a}{\sigma x}) \ge \log(1 + \frac{a}{2n}) \ge \frac{a}{2n}$ (we used the inequality $\log(1 + z) \ge z$, where $0 \le z \le 1$). Thus, the redundancy is approximately $\frac{a}{2n}$ bits per symbol $a$, which sums to the total redundancy of $\sum_a \frac{a}{2n}f_a = \sum_a \frac{a}{2\sigma} = \frac{\sigma-1}{4}$ bits over all symbols in the sequence.

This informal argument is only an intuition since the non-positive terms $\log\left(\frac{1 + \Delta / (x' 2^\ell)}{1 + \Delta / (x 2^\ell)}\right)$ that appear in the analysis of Section~\ref{sec:analysis-upper-bound} could, in principle, diminish the described effect. Nevertheless, as we are to show, an~$\Omega(\sigma)$ redundancy indeed appears in some instances.

\medskip

Fix an even integer $r > 0$. Observe that $2^r \equiv 1 \pmod{3}$ since $r$ is even. Denote $n = 2^r$. The sequence under construction will contain $\sigma = (n - 1) / 3 + 1$ symbols $0, 1, \ldots, \sigma - 1$. Each symbol $a \in [0..\sigma{-}1)$ has exactly three occurrences in the sequence (i.e., $f_a = 3$) and the symbol $\sigma - 1$ occurs only once (i.e., $f_{\sigma-1} = 1$); note that $\sum_{a\in [0..\sigma)} f_a = 3(\sigma - 1) + 1 = n$. The entropy formula gives the following lower bound on the encoding size for the sequence:
\begin{equation}
(n - 1)\log\frac{n}{3} + \log n = (n - 1)(r - \log 3) + r = (n - 1)(r - 1.58496...) + r.\label{eq:example-opt}
\end{equation}
It is straightforward that with such frequencies of symbols both Duda's initialization algorithm and its simplified variant (Algorithm~\ref{alg:init}) construct the same array $\mathsf{range}$: the subrange $\mathsf{range}[0..\sigma{-}1]$ contains consecutively the symbols $0, 1, \ldots, \sigma - 1$ (in this order) and the subranges $\mathsf{range}[\sigma..2\sigma{-}2]$ and $\mathsf{range}[2\sigma{-}1..n{-}1]$ are equal and both contain consecutively the symbols $0, 1, \ldots, \sigma - 2$ (in this order). Now let us arrange the symbols in the sequence $a_1, a_2, \ldots, a_n$.

The last symbol $a_n$ is $\sigma - 1$. The rest, $a_1, a_2, \ldots, a_{n-1}$, consists of symbols $a \in [0..\sigma{-}1)$ whose frequencies are $f_a = 3$ ($11_2$ in binary). When the encoder processes a symbol $a_{i+1} = a \in [0..\sigma{-}1)$ and modifies the number $w_i$ representing the prefix $a_1, a_2, \ldots, a_i$, it replaces either two or three leading bits of $w_i$ with new $r+1$ bits, thus producing the number $w_{i+1}$. The choice of whether to replace two or three bits depends on whether the two leading bits of $w_i$ are $11$ or $10$, respectively (i.e., whether the two bits store a number less than $f_a = 3$ or not). We are to arrange the symbols $[0..\sigma{-}1)$ in the sequence $a_1, a_2, \ldots, a_{n-1}$ in such a way that the encoder chooses the two options alternatingly: it replaces three leading bits of $w_i$ if $i$ is even, and two bits if $i$ is odd ($i = 0,1,\ldots,n-2$). The total number of bits produced in this way is at least $\frac{n}{2}(r - 2) + (\frac{n}{2} - 1)(r - 1) + r$ (the additive term $r$ appears when the last symbol $a_n = \sigma - 1$ is encoded), which is equal to $(n - 1)(r - 1.5) + r - 0.5$. Comparing this to~\eqref{eq:example-opt}, one can see that the encoding generated by ANS is larger than the optimum~\eqref{eq:example-opt} by at least $(\log 3 - 1.5)(n-1) - 0.5 > 0.08496 (n-1)$ (the estimation holds for large enough $n$, so that the term $0.5$ disappears). By simple calculations, we deduce from the equality $\sigma = (n - 1) / 3 + 1$ that the redundancy $0.08496 (n-1)$ is larger than $\frac{\sigma-1}{4}$. Let us describe an arrangement of symbols that produces such effect of ``alternation''.

The encoder consecutively transforms the initial value $w_0 = 2^{r}$ into $w_1, w_2, \ldots$ by performing the push operations: $w_{i+1} = \mathsf{push}(w_i, a_{i+1})$. Let us call  the number $\lfloor w_i / 2^{\lfloor\log w_i\rfloor - r}\rfloor \bmod 2^r$ a \emph{state}; it is the value stored in the highest $r+1$ bits of the number $w_i$ currently processed by the encoder minus the highest bit~1.
The range of possible states is $[0..2^r) = [0..n)$. We split this range into three disjoint segments numbered i, ii, iii: $[0..\sigma)$, $[\sigma..2\sigma - 1)$, $[2\sigma - 1..n)$, whose lengths are $\sigma$, $\sigma - 1$, $\sigma - 1$, respectively. The subarray $\mathsf{range}[0..\sigma{-}1]$ corresponding to the segment~i contains all symbols $[0..\sigma)$; each of the subarrays $\mathsf{range}[\sigma..2\sigma - 2]$ and  $\mathsf{range}[2\sigma - 1..n - 1]$ corresponding to the segments~ii and~iii, respectively, contains all symbols $[0..\sigma{-}1)$.

\begin{figure*}[ht]
\centering
\includegraphics{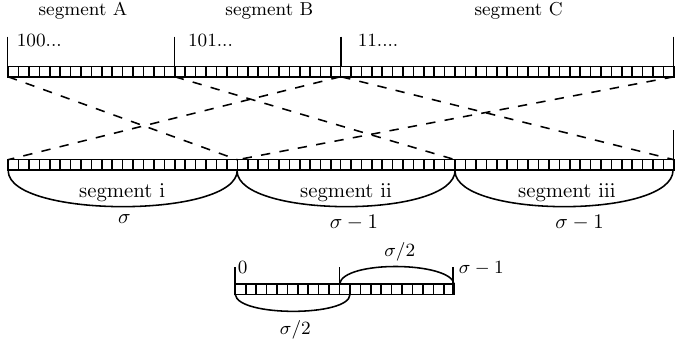}
\caption{Two splits of the state range $[0..2^r)$ into segments. Here $r = 6$. Common first bits of numbers $w_i$ corresponding to the states belonging to the segments A, B, C are written above the respective segments. Observe that the bits start with $11$ only when a state belongs to the segment C. The range $[0..\sigma{-}1)$ is depicted under the segment~ii: the encoder receives symbols from the right part $[\sigma/2 - 1..\sigma - 1)$ when the current state is from the segment~A, and from the left part $[0..\sigma/2)$ when the state is from the segment~C.}\label{fig:transits}
\end{figure*}

Receiving the symbol $\sigma - 1$ (which occurs in the sequence only once), the encoder transforms any current state to the state $\sigma - 1$ (it is the index of the symbol $\sigma - 1$ in the array $\mathsf{range}$). In order to describe how states are transformed by other symbols, let us split the state range $[0..2^r)$ into three segments called A, B, C: $[0..2^{r-2})$,  $[2^{r-2}..2^{r-1})$,  $[2^{r-1}..n)$, whose lengths are $2^{r-2}$, $2^{r-2}$, $2^{r-1}$, respectively. The segments are related as follows: receiving a symbol from $[0..\sigma{-}1)$, the encoder translates the current state from the segment A (respectively, B, C) to a state from the segment~ii (respectively, iii,~i); see Figure~\ref{fig:transits} where this transition of states is illustrated by dashed lines connecting the corresponding segments.

Receiving $a_{i+1} = a \in [0..\sigma{-}1)$, the encoder replaces two leading bits of $w_i$ with $r + 1$ new bits if the current state is from the segment~C; otherwise, it replaces three leading bits of $w_i$. This behaviour is determined by first bits of $w_i$, which are equal to $11...$ only for states from the segment C; see Figure~\ref{fig:transits}. The execution of the encoder starts with the state $0$, which belongs to the segment~A. We are to arrange symbols in the sequence $a_1, a_2, \ldots, a_{n-1}$ so that, for odd $i$, the state corresponding to $w_i$ belongs to the segment~C, and for even $i$, to the segment~A. The states in our arrangement will never belong to the segment~B.

Since $\sigma + \sigma/2 - 1 = n/2$ and $n/2$ is the leftmost state from the segment~C, any state from the segment~A transits to a state from the segment~C when the encoder receives a symbol $a$ from the range $[\sigma/2 - 1..\sigma - 1)$; the new state is $\sigma + a$, the $a$th element of the segment~ii (see Figure~\ref{fig:transits} for an illustration). Similarly, since $\sigma / 2 < n/4$ and $n/4 - 1$ is the rightmost state from the segment~A, any state from the segment~C transits to a state from the segment~A when the encoder receives a symbol $a \in [0..\sigma/2)$. Accordingly, for $i \in [0..n{-}1)$, we put in the sequence as the symbol $a_{i+1}$ a symbol from $[\sigma/2 - 1..\sigma - 1)$ if $i$ is even, and a symbol from $[0..\sigma/2)$ if $i$ is odd (note that both ranges share a common symbol $\sigma/2-1$; it is not a mistake). The states thus ``bounce'' between the segments~A and~C during the processing of $a_1, a_2, \ldots, a_{n-1}$ by the encoder. Since the sizes of both ranges $[0..\sigma/2)$ and $[\sigma/2 - 1..\sigma - 1)$ are $\sigma/2$ and they share a common symbol $\sigma/2-1$, the symbols $[0..\sigma{-}1)$ can be distributed in the sequence $a_1, a_2, \ldots, a_{n-1}$ in such way that each symbol occurs exactly three times and each symbol $a_{i+1}$ is from the range $[\sigma/2 - 1..\sigma - 1)$, for even $i$, and from the range $[0..\sigma/2)$, for odd $i$.

We thus have obtained a redundancy of $\frac{\sigma - 1}{4}$ bits. Adding to this $r - 2$ bits produced by the $r-2$ lowest bits of the initial value $w_0$, we have proved the following theorem.

\begin{theorem}
For arbitrarily large $n = 2^r$ with even $r$, there exists an alphabet $[0..\sigma)$, where $\sigma > n/3$, and a sequence $a_1, a_2, \ldots, a_n$ of symbols from this alphabet such that the ANS encoder using [simplified] Duda's precise initialization transforms this sequence into a bit string of length at least
$$
\sum\limits_{a \in [0..\sigma)} f_a \cdot \log\frac{n}{f_a} + \frac{\sigma - 1}{4} + r - 2,
$$
where $f_a$  is the number of occurrences for symbol $a$. Thus, the redundancy is $\frac{\sigma - 1}{4} + r - 2 = \Theta(n)$ bits.
\label{thm:lower-bound}
\end{theorem}

Note that Theorem~\ref{thm:lower-bound} does not account for the space required to encode the table of symbol frequencies (which itself will be very large when $\sigma > n/3$) since, as it was stated in the preliminary section, the problem of storage for the frequency table is out of the scope of the present paper.

The example that attains the lower bound of Theorem~\ref{thm:lower-bound} is simple (perhaps, unlike its tedious analysis). Hence, it is reasonable to assume that any adequate shuffling method would have the same redundancy $\Omega(\sigma + r)$ as in Theorem~\ref{thm:lower-bound} on a similarly constructed sequence $a_1, a_2, \ldots, a_n$. We believe, therefore, that Duda's conjecture that the redundancy can be $O(\frac{\sigma}{n^2})$ bits per symbol when an appropriate shuffling method is used is disproved. This probably is not the case for algorithms that construct the shuffling tables after scanning the sequence first. However, such methods seem infeasible in practice due to incurring performance losses.

\section{Range ANS with Fixed Accuracy}\label{sec:enhanced-rans}

In Section~\ref{sec:rans}, a simple range ANS (rANS) was described, which is just the ANS without shuffling. Now we are to introduce another variant of rANS, called \emph{rANS with fixed accuracy} to distinguish it from the rANS as defined by Duda~\cite{Duda2013,DudaEtAl} (which is also sketched below). Our exposition is less detailed than in the previous sections since we believe that all ideas and intuition necessary for understanding were developed above.

Let $a_1, a_2, \ldots, a_n$ be a sequence of symbols over an alphabet $[0..\sigma)$, where $n = 2^r$ and the frequencies of symbols are denoted by $f_a$ (i.e., the empirical probability for symbol $a$ is $\frac{f_a}{n}$). Fix an integer $k \ge 0$, which will serve as a user-defined accuracy parameter regulating the size of redundancy (typically, $0 \le k \le 4$).
As in the simple rANS, the encoder starts its work with a number $w_0 = 2^{r+k}$; the value $w_0$ might be arbitrary, the only necessary condition is $w_0 \ge 2^{r+k}$. The encoder consecutively performs operations $w_{i+1} = \mathsf{push}(w_i, a_{i+1})$, for $i = 0, 1,\ldots, n-1$, but the operation $\mathsf{push}(w, a)$ works differently this time: it substitutes either $\lfloor\log f_a\rfloor + k + 1$ or $\lfloor\log f_a\rfloor + k + 2$ highest bits of $w$ that store a number $x$ by $r + k + 1$ new bits that store the number $x' = \lfloor x / f_a \rfloor 2^r + c_a + (x \bmod f_a)$, where $c_a = \sum_{b \in [0..a)} f_b$; the condition determining the number of bits occupied by $x$ is essentially the same as in the simple rANS from Section~\ref{sec:rans}.

More formally, the algorithm for $\mathsf{push}(w, a)$ is as follows. Denote by $x_1$ and $x_2$ the values stored in, respectively, the highest $\lfloor\log f_a\rfloor + k + 1$ and $\lfloor\log f_a\rfloor + k + 2$  bits of $w$. We assume that $x = x_1$ if $x_1 \ge f_a 2^k$, and $x = x_2$ otherwise. Note that $2^{\lfloor\log f_a\rfloor + k} \le x_1 < f_a 2^{k+1}$ and $f_a 2^k < 2^{\lfloor\log f_a\rfloor + k + 1} \le x_2 \le 2\cdot x_1 + 1$. If $x = x_1$ (i.e., $f_a 2^k \le x_1$), we have $f_a 2^k \le x < f_a 2^{k+1}$. If $x = x_2$ (i.e., $f_a 2^k > x_1$), we have $x \le 2\cdot x_1 + 1 < f_a 2^{k+1}$. Thus, both cases imply $f_a 2^k \le x < f_a 2^{k+1}$. Denote $\ell = \lfloor\log w\rfloor - \lfloor\log x\rfloor$. Note that $w = x \cdot 2^{\ell} + (w \bmod 2^\ell)$. The value $w' = \mathsf{push}(w, a)$ is computed as $w' = x' \cdot 2^{\ell} + (w \bmod 2^{\ell})$ by replacing the part $x$ of $w$ by $r + k + 1$ bits representing a number $x'$ defined as:
\begin{equation}
x' = \lfloor x / f_a\rfloor 2^r + c_a + (x \bmod f_a).\label{eq:x-prime-rans}
\end{equation}
Since $f_a 2^k \le x < f_a 2^{k+1}$, we have $2^k \le \lfloor x / f_a\rfloor < 2^{k+1}$ and, therefore, the number $x'$ indeed fits into $r + k + 1$ bits and its highest $(r + k + 1)$th bit is~1.
The reverse operation $(w,a) = \mathsf{pop}(w')$ producing the old value $w$ and the symbol $a$ from $w'$ is straightforward. Given a number $x'$, which occupies $r + k + 1$ highest bits of $w'$ (i.e., $x' = \lfloor w' / 2^{\ell}\rfloor$, where $\ell = \lfloor \log w'\rfloor -r - k$), we first determine the symbol $a$ by examining to which range $[c_a..c_{a+1})$ the number $x' \bmod 2^r$ belongs and, then, we compute $x$ as follows:
\begin{equation}
x = \lfloor x' / 2^r\rfloor f_a + (x' \bmod 2^r) - c_a.\label{eq:x-rans}
\end{equation}
Once $x$ is known, we put $w = x\cdot 2^{\ell} + (w' \bmod 2^{\ell})$. It remains to observe that $x' > x$ since $x' = \lfloor x / f_a\rfloor 2^r + (x \bmod f_a) > \lfloor x / f_a\rfloor f_a + (x \bmod f_a) = x$. Therefore, we have $w_0 < w_1 < \cdots < w_n$.

Note that, when $k = 0$, the described scheme degenerates simply to the ANS without shuffling.

To estimate the size of the final number $w_n$ in bits, we derive by analogy to~\eqref{eq:wn} and~\eqref{eq:w-to-x} using the condition $x' > x$ the following two equations (here we have $w_i = x\cdot 2^\ell + \Delta$ and $w_{i+1} = x'\cdot 2^\ell + \Delta$, where $\Delta \in [0..2^\ell)$):
$$%\begin{equation}
\log w_n = \log\frac{w_n}{w_{n-1}} + \log\frac{w_{n-1}}{w_{n-2}} + \cdots + \log\frac{w_1}{w_0} + r + k; %\tag{\ref{eq:wn}}
$$%\end{equation}
$$%\begin{equation}
\log w_{i+1} - \log w_i \le \log x' - \log x. %\tag{\ref{eq:w-to-x}}
$$%\end{equation}
From~\eqref{eq:x-rans}, we deduce $\log x \ge \log(\lfloor x' / 2^r\rfloor f_a) \ge \log f_a + \log (x' / 2^r - 1) = \log f_a + \log(x' / 2^r) + \log(1 - 2^r / x') = \log \frac{f_a}{2^r} + \log x' + \log(1 - 2^r / x')$. Since $2^{r+k} \le x'$, we derive further $\log(1 - 2^r / x') \ge \log(1 - 1 / 2^k) \ge -\frac{\log e}{2^k - 1}$ (due to the inequality $\ln(1 - z) \ge \frac{-z}{1-z}$, for $0 \le z < 1$). Therefore, we obtain
$$
\log x' - \log x \le \log\frac{n}{f_a} + \frac{\log e}{2^k - 1}.
$$

Now we derive  $\log w_n = \sum_{i=0}^{n-1} (\log w_{i+1} - \log w_i) + r + k \le \sum_{i=0}^{n-1} (\log\frac{n}{f_{a_{i+1}}} + \frac{\log e}{2^k-1}) + r + k = \sum_{a\in [0..\sigma)} f_a\cdot \log\frac{n}{f_a} + \frac{n\log e}{2^k-1} + r + k$, getting the following theorem.

\begin{theorem}
Given a sequence $a_1, a_2, \ldots, a_n$ of symbols from an alphabet $[0..\sigma)$ such that each symbol $a$ occurs in it $f_a$ times and $n = 2^r$ for an integer $r$, the rANS encoder with fixed accuracy $k \ge 1$ transforms this sequence into a bit string of length
$$
\sum\limits_{a \in [0..\sigma)} f_a \cdot \log\frac{n}{f_a} + O\left(\frac{n}{2^k} + r + k\right),
$$
where the $O\left(\frac{n}{2^k} + r + k\right)$ redundancy term can be bounded by $\frac{n\log e}{2^k - 1} + r + k$.\label{thm:upper-bound-rans}
\end{theorem}

For completeness, let us briefly describe the standard rANS~\cite{Duda2013,DudaEtAl}. The encoder similarly starts with the number $w_0 = 2^r$ and consecutively computes $w_1, w_2, \ldots, w_n$ for the sequence $a_1, a_2, \ldots, a_n$, where $n = 2^r$. The encoder maintains a number $h$ (initially $h = r + 1$) and, receiving a new symbol, it replaces the highest $h$ bits of the current number $w_i$ with a larger value and increases $h$ accordingly. A ``renormalization'' is sometimes performed by reducing the value $h$ in order to contain numbers within a range fitting into a machine word. More precisely, receiving a symbol $a = a_{i+1}$, the encoder takes the value $x$ stored in the $h$ highest bits of $w_{i}$ (i.e., $w_i = x\cdot 2^\ell + \Delta$, where $\ell = \lfloor\log w_i\rfloor + 1 - h$ and $\Delta \in [0..2^\ell)$), calculates a number $x'$ by formula~\eqref{eq:x-prime-rans}, and replaces $x$ with $x'$, thus producing $w_{i+1} = x'\cdot 2^\ell + \Delta$. After this, the value $h$ is increased by $\lfloor\log x'\rfloor - \lfloor\log x\rfloor$. Once $h$ is larger than a fixed threshold $H$, it is decreased but the resulting $h$ should be larger than $r$. Usually, for performance reasons, encoders decrease $h$ by a multiple of 8 or 16: $h = r + 1 + ((h - r - 1) \bmod b)$, where $b = 8$ or $b = 16$. Implementations maintain the $h$-bit number $x$, assigning $x = x'$ after processing each symbol, and the decreased number of bytes from $x$ are ``dumped'' into an external stream. The decoder executes the same operations but in the reverse order computing $x$ from $x'$ as in~\eqref{eq:x-rans}.

The analysis of this rANS variant is not in the scope of the present paper; see~\cite{Duda2013,YamamotoIwata,Townsend}.

\paragraph{Implementation notes}
The code of our implementation and tests is available at \url{https://github.com/dkosolobov/rans_with_fixed_accuracy}. See Table~\ref{tbl:experiments}.
We tested it against the standard rANS with explicit divisions and against a faster rANS with divisions performed by multiplications and shifts on precomputed constants; these two rANS variants were retrieved from the source code of \texttt{ryg\_rans} by Giesen.\footnote{\url{https://github.com/rygorous/ryg_rans}} 
We note that neither of the four tested encoders implements the so-called interleaving streams: this technique can noticeably speed up the rANS~\cite{Giesen} but it will do so equally for all four variants, if implemented, so we decided to omit it since, on a qualitative level, the results will likely remain the same with or without this technique.

All experiments were conducted on a laptop with AMD Ryzen~7 5800H (3.2\,GHz, 512\,Kb L1 cache, 16\,Mb L3 cache) with 16\,Gb RAM under 64-bit Windows 11 operating system. The code was compiled in Visual Studio 2019 using the {\tt Clang} compiler; the essential flags of compilation were {\tt-O2} {\tt-Oi} {\tt-Ot} {\tt-Gd} {\tt-GS} {\tt-arch:AVX2} (plus some technical flags included by the Visual Studio). The time was measured as average over five runs.

In our experiments the described rANS encoders with fixed accuracies $k=3$ and $k=2$ had approximately the same compression rate as the standard rANS (with a slight noticeable overhead for $k = 2$). In terms of speed, the rANS with accuracies $k=3$ and $k = 2$ provide a faster encoding than the classical rANS with divisions but slower than the faster rANS implementation; however, the variant with $k = 2$ is very close to the faster rANS implementation. The rANS with accuracies $k=3$ and $k=2$ both were consistently slower in decoding.

The key feature that allows the speed boost in encoding is that the operation of division $\lfloor x / f_a\rfloor$ in~\eqref{eq:x-prime-rans} guarantees that its result is in the range $[2^k..2^{k+1})$. Therefore, the division can be executed by simpler instructions in a branchless code: we used arithmetic and bit operations and the instruction \texttt{cmov} from x86 (it is also possible to use only arithmetic and bit instructions). The code, however, turns out to be quite cumbersome, which noticeably diminishes positive effects of the division-free branchless loop.

\begin{table}
	\caption{Four rANS variants on three inputs. Each input has length 65536 with byte alphabet and it is generated either by a geometric or uniform distribution. The right column shows the compressed size (without frequency tables). For details, see \texttt{https://github.com/dkosolobov/rans\_with\_fixed\_accuracy}.}
%\begin{tabular}{l|l|l|l}
%	Input &	Entropy encoder &	Comp/decomp time & Out bytes \\\hline
%	Geom. dist. $p = 0.7$ &	rANS with acc 3: &	241500/408800 ns &	10366\\
%	~ & rANS: &	305700/285500 ns &	10364\\
%	~ & rANS fast: &	182500/284400 ns &	10364\\\hline
%	Geom. dist. $p = 0.3$ &	rANS with acc 3: &	235600/407200 ns &	24117\\
%	~ & rANS: &	292900/275600 ns &	24112\\
%	~ & rANS fast: &	185700/276900 ns &	24112\\\hline
%	Uniform dist. &	rANS with acc 3: &	235500/398400 ns &	65521\\
%	~ & rANS: &	257000/256000 ns &	65516\\
%	~ & rANS fast: &	152200/256600 ns &	65516
%\end{tabular}\label{tbl:experiments}
%
\begin{tabular}{l|l|l|l}
	Input &	Entropy encoder &	Comp/decomp time & Out bytes \\\hline
Geom. dist. $p = 0.7$ &	rANS with acc 3: &	341780/412960 ns &	10366\\
~                     & rANS with acc 2: &	288740/399640 ns &	10372\\
~                     & rANS: 			 &	403560/286800 ns &	10364\\
~                     & rANS fast: 		 &	278200/286560 ns &	10364\\\hline
Geom. dist. $p = 0.3$ &	rANS with acc 3: &	270120/391740 ns &	24117\\
~ 					  & rANS with acc 2: &	237240/393480 ns &	24135\\
~ 					  & rANS: 			 &	317800/259460 ns &	24112\\
~ 					  & rANS fast: 		 &	218480/261520 ns &	24112\\\hline
Uniform dist. 		  &	rANS with acc 3: &	266920/396760 ns &	65521\\
~ 					  & rANS with acc 2: &	231240/392520 ns &	65545\\
~ 					  & rANS: 			 &	312480/252580 ns &	65516\\
~ 					  & rANS fast: 		 &	179380/253480 ns &	65516
\end{tabular}\label{tbl:experiments}
\end{table}

Denote $R = r + k$. The main loop of the encoder calls the function $\mathsf{encode}$ from Algorithm~\ref{alg:rans} consecutively for the symbols $a_1, a_2, \ldots, a_n$ in the encoded sequence. The function receives as its parameters an $(R+1)$-bit number $w$ and a symbol $a$. The function stores some lowest bits of $w$ in an external storage and returns an $(R+1)$-bit value $x'$ computed as in~\eqref{eq:x-prime-rans} (details follow). The parameter $w$ is actually an $(R+1)$-bit number $x'$ produced by the previous call to the function $\mathsf{encode}$ in the encoding loop; the first call receives $w = 2^R$.

\begin{algorithm}
\caption{The encoding function of rANS with fixed accuracy $k = 3$.}\label{alg:rans}
\begin{algorithmic}[1]
\Function{$\mathsf{encode}$}{$w, a$} \Comment{rANS with fixed accuracy $k = 3$}
\State $(c_a, f_a, d) = \mathsf{table}[a];$ \Comment{$d = (t \shl (R + 1)) - (f_a \shl (t + k))$, where $t = r - \lfloor \log f_a\rfloor$}
\State $s = (w + d) \shr (R + 1);$            \Comment{Collet's trick}
\State $\mathsf{outBits}(w, s);$   \Comment{output $s$ lowest bits of $w$}
\State $x = w \shr s;\;$
\State $x = x - (f_a \shl 3);$\label{lst:div-beg}
\State $q = 0;$
\For{$i = 2,1,0$} \Comment{the loop must be unrolled}
    \State $x_0 = x - (f_a \shl i);\;$
    \State $\mathbf{if}\, (x_0 \ge 0)\; x = x_0;\;$  \Comment{compiled to \texttt{cmov} on x86}
    \State $q = q  \mathrel{\mathbf{or}}  (x_0  \mathrel{\mathbf{and}} (1 \shl (R + i)));$\label{lst:div-end}
\EndFor
\iffalse
\State $x_0 = x - (f_a \shl 2);\;$
\State $\mathbf{if}\, (x_0 \ge 0)\; x = x_0;\;$
\State $q = x_0  \mathrel{\mathbf{and}}  (4 \shl R);\;$
\State $x_0 = x - (f_a \shl 1);\;$
\State $\mathbf{if}\, (x_0 \ge 0)\; x = x_0;\;$
\State $q = q \mathrel{\mathbf{or}} (x_0 \mathrel{\mathbf{and}} (2 \shl R));$
\State $x_0 = x - f_a;\;$
\State $\mathbf{if}\, (x_0 \ge 0)\; x = x_0;\;$
\State $q = q  \mathrel{\mathbf{or}}  (x_0  \mathrel{\mathbf{and}} (1 \shl R));$
\fi%\label{lst:div-end}
\State\Return{$((q \mathrel{\mathbf{xor}} (15 \shl R)) \shr k) + c_a + x;$}
\EndFunction
\end{algorithmic}
\end{algorithm}

Denote $t = r - \lfloor \log f_a\rfloor$. The number $x$ occupies either $R - t + 1$ or $R - t + 2$ highest bits of $w$. The presented pseudocode uses Collet's trick~\cite{Collet} to determine $x$ with a branchless code. To this end, the array $\mathsf{table}$ stores, for each symbol $a \in [0..\sigma)$, besides the values $c_a$ and $f_a$ the number $d = (t \shl (R + 1)) - (f_a \shl (t + k))$. The trick is that, in this case, the number $w + d = (t \shl (R + 1)) + w - (f_a \shl (t + k))$ contains in its highest bits $R + 1, R + 2, \ldots$ either the number $t$ or $t - 1$ depending on whether $w \ge (f_a \shl (t + k))$ or not. Therefore, $x = w \shr s$, where $s = (w + d) \shr (R + 1)$.

The code in lines~\ref{lst:div-beg}--\ref{lst:div-end} accumulates the quotient $\lfloor x / f_a\rfloor$ in bits $R+1,R+2,\ldots$ of the variable $q$ and the remainder $x \bmod f_a$ in the variable $x$. It is done by subtracting the numbers $f_a \shl i$, for $i = 3,2,1,0$, from $x$, thus, reconstructing $q$ bit by bit; note, however, that the bits in $q$ are inverted and, hence, in the end we have to perform xor with $15$ ($1111_2$ in binary).

As is usual for such encoders, the function $\mathsf{outBits}(w,s)$ outputs $s$ lower bits of $w$ into a separate ``buffer'' machine word; this machine word is ``dumped'' into an external array in chunks of size 8 bits when too many bits are stored in it; see the details in \url{https://github.com/dkosolobov/rans_with_fixed_accuracy}.

As for the decoding process, it does not differ significantly from the decoding in the standard rANS and is quite straightforward; see the details in  \url{https://github.com/dkosolobov/rans_with_fixed_accuracy}. However, the decoding has more instructions and, hence, is slightly slower.

\paragraph{Discussion.}
The standard rANS has a couple of advantages over tANS in some use cases~\cite{Giesen}: it does not require a table of size $2^r$ like tANS and, hence, is more convenient for the (pseudo) adaptive mode when the table of frequencies is sometimes rebuilt during the execution of the encoder; due to this less heavy use of memory, the rANS might be better for interleaving several streams of data and utilizes more efficiently the instruction-level parallelism for this task. For these reasons, the rANS was used in some high performance compressors~\cite{Giesen,TurboRC}. 
We note that in the (pseudo)adaptive setting the precomputation of constants for fast division is expensive and, hence, the standard rANS with explicit division is often used in this scenario. The described rANS with fixed accuracy shares the same good features of the standard rANS plus the described above benefits of the controlled division (at the expense of slower decoding, however). 

An accurate integration of the rANS with fixed accuracy into existing adaptive encoders is an engineering task that we leave for a future work. One might expect that the speed up will be negligible if the time in the adaptive case is dominated by the updates of cumulative frequency tables. However, the (pseudo)adaptive setting is typically applied when the alphabet size is relatively small (at most 16), which allows to use SIMD instructions for fast updates of the cumulative frequencies, thus making the adaptive encoding not dramatically slower than the non-adaptive encoding; e.g., see the TurboRC encoder in \cite{TurboRC} and benchmarks therein. Still, in view of the above experimental results, only a minor speed up might be expected.

\section{Conclusion and Open Problems}\label{sec:conclusion}

Theorems~\ref{thm:upper-bound} and~\ref{thm:lower-bound} describe the tight asymptotic behaviour of the redundancy for the tANS. We believe that Theorem~\ref{thm:upper-bound-rans}, albeit not complemented with a lower bound, is asymptotically tight too. However, it is open to provided a series of examples supporting this claim. A number of other problems listed below still remain open too.

(i) The main remaining open problem concerning ANS encoders, as we see it, is to construct a genuine FIFO encoder with the same performance characteristics as tANS or rANS. The known ANS variants work as stacks (LIFO) while it is more natural and, in some scenarios, preferable to have an encoder that acts as a queue, like the arithmetic coding that fulfils this requirement but is noticeably worse than ANS in performance terms. For the same reason, the current ANS variants are not suitable enough for the adaptive encoding when frequencies of symbols change as the algorithm reads the sequence from left to right. 

We note that the rANS can be applied for the adaptive mode as follows (for instance, see~\cite{TurboRC}): it first performs one left-to-right pass on the sequence to collect individual statistics per symbol adaptively and, then, encodes the sequence from right to left using these statistics; the decoder decodes the sequence from left to right adaptively computing the statistics by the same algorithm as was used by the encoder in the first pass. At first glance, this scheme is not much different from what the tANS or non-adaptive rANS do (note that they also perform a separate pass collecting statistics). Unfortunately, the adaptive rANS cannot utilize many optimizations that make a modern rANS comparable to tANS in terms of speed (in particular, the decoder cannot use SIMD and interleaving streams \cite{Giesen} as freely and cannot speed up divisions with special precomputed values). One of the goals for a FIFO encoder is to fix this issue.

(ii) Duda's precise initialization and its simplified variant from Algorithm~\ref{alg:init} are not particularly suitable for practice due to the overhead incurred by operations on the priority queue and by floating point operations. Therefore, usually they are replaced with heuristics. An implementation of a fast and simple initialization method with good guarantees sufficient for Theorem~\ref{thm:upper-bound} is an open problem.

(iii) The constants in our lower and upper bounds in Theorems~\ref{thm:upper-bound} and~\ref{thm:lower-bound} do not coincide and it remains open to find a tight constant for the tANS redundancy term. We believe also that the $r$-bit redundancy produced by the initial state $w_0$ of the encoder can be somehow reduced too by slightly modifying the scheme. Clearly, one can omit the trailing zeros in the bit representation of the resulting number $w_n$ but it does not suffice to get rid of the $r$-bit redundancy entirely.

(iv) The encoding of the table of frequencies is rarely in the focus of theoretical research. However, it is very important in practice, mainly due to the following observation: a typical sequence fed to the entropy encoder is not homogeneous but consists of ``chunks'' drawn from different distributions; accordingly, an optimized compressor normally either uses an adaptive encoder (which seems best for this case but such encoders currently have issues described above) or splits the sequence into blocks and encodes each block separately, each with its own frequency table (e.g., see~\cite{AlakuijalaBrotli}). On relatively small chunks of data such as those produced in the latter case, the size of the frequency table is not negligible compared to the entropy of the data. For this reason, many practitioners resort to simpler Huffman encoders, which require much less space to encode their frequency tables. Surprisingly, this advantage in practice often suffices to compensate for the larger redundancy of Huffman compared to arithmetic or ANS encoders. Note that it is not an exceptional case: for instance, the Huffman encoder is \emph{consistently superior} to arithmetic or ANS when encoding literals in LZ77 compressors (some LZ77 compressors often cautiously utilize the Huffman encoders in other parts as well, like to encode LZ77 phrase offsets and phrase lengths, without losses in the compression ratio; however, in most cases these parts still are processed by ANS or arithmetic encoders). Often the authors of compressors resort to the Huffman encoders after extensive experiments with ANS and arithmetic encoders, so the reason is not in simplicity.

Theoretical tight bounds are known for the size of encodings of the frequency tables: see \cite{Shtarkov}, \cite{OrlitskySanthanam}, \cite{SzpankowskiWeinberger}, and references therein. However, works that investigate practical encoders for these tables are scarce. Ideally, such an encoder should take into account the entropy of the data to be able to relax some frequencies in order to make the total encoding size smaller, if necessary, thus never being worse than the Huffman encoders (see~\cite{Duda2021} and references in~\cite{PainskyRossetFeder}). It seems that further investigations in this direction are needed.

\section*{Acknowledgments}

I wish to thank the anonymous referees whose detailed reviews and suggestions helped to improve the paper (in particular, for pointing out that a recent result of \cite{YamamotoIwata2} follows from Theorem~\ref{thm:upper-bound2}).

\section*{Declarations}

This work was supported by the Russian Science Foundation (RSF), project 24-71-00062.
The author has no conflicts of interest to declare that are relevant to the content of this article.

%\section*{References}
%\bibliographystyle{elsart-num-sort}
%\bibliographystyle{IEEEtran}
\bibliographystyle{unsrt}
\bibliography{refs}

\begin{thebibliography}{10}

\bibitem{Duda2009}
J.~Duda.
\newblock Asymmetric numeral systems.
\newblock {\em arXiv preprint arXiv:0902.0271}, pages 1--47, 2009.

\bibitem{Duda2013}
J.~Duda.
\newblock Asymmetric numeral systems: entropy coding combining speed of huffman
  coding with compression rate of arithmetic coding.
\newblock {\em arXiv preprint arXiv:1311.2540}, pages 1--24, 2013.

\bibitem{DudaEtAl}
J.~Duda, K.~Tahboub, N.~J. Gadgil, and E.~J. Delp.
\newblock The use of asymmetric numeral systems as an accurate replacement for
  {H}uffman coding.
\newblock In {\em 2015 Picture Coding Symposium (PCS 2015)}, pages 65--69.
  IEEE, 2015.

\bibitem{Martin}
G.~N.~N. Martin.
\newblock Range encoding: an algorithm for removing redundancy from a digitised
  message.
\newblock In {\em Proc. Institution of Electronic and Radio Engineers
  International Conference on Video and Data Recording}, page~48, 1979.

\bibitem{MoffatNealWitten}
A.~Moffat, R.~M. Neal, and I.~H. Witten.
\newblock Arithmetic coding revisited.
\newblock {\em ACM Transactions on Information Systems}, 16(3):256--294, 1998.

\bibitem{Pasco}
R.~C. Pasco.
\newblock {\em Source coding algorithms for fast data compression}.
\newblock PhD thesis, Stanford University CA, 1976.

\bibitem{Rissanen}
J.~J. Rissanen.
\newblock Generalized {K}raft inequality and arithmetic coding.
\newblock {\em IBM Journal of research and development}, 20(3):198--203, 1976.

\bibitem{WittenNealCleary}
I.~H. Witten, R.~M. Neal, and J.~G. Cleary.
\newblock Arithmetic coding for data compression.
\newblock {\em Communications of the ACM}, 30(6):520--540, 1987.

\bibitem{Huffman}
D.~A. Huffman.
\newblock A method for the construction of minimum-redundancy codes.
\newblock {\em Proc. Institute of Radio Engineers (IRE)}, 40(9):1098--1101,
  1952.

\bibitem{Collet}
Y.~Collet.
\newblock {FSE implementation}.
\newblock \url{https://github.com/Cyan4973/FiniteStateEntropy}.
\newblock Accessed: 07.01.2022.

\bibitem{LZFSE}
{Apple LZFSE compressor}.
\newblock \url{https://github.com/lzfse/lzfse}.
\newblock Accessed: 07.01.2022.

\bibitem{ColletZstd}
Y.~Collet.
\newblock {Facebook Zstandard compressor}.
\newblock \url{https://github.com/facebook/zstd}.
\newblock Accessed: 07.01.2022.

\bibitem{Giesen}
F.~Giesen.
\newblock Interleaved entropy coders.
\newblock {\em arXiv preprint arXiv:1402.3392}, pages 1--16, 2014.

\bibitem{AlakuijalaEtAl}
J.~Alakuijala, R.~van Asseldonk, S.~Boukortt, M.~Bruse, I-M. Com{\c{s}}a,
  M.~Firsching, T.~Fischbacher, E.~Kliuchnikov, S.~Gomez, R.~Obryk, et~al.
\newblock {JPEG XL} next-generation image compression architecture and coding
  tools.
\newblock In {\em Applications of Digital Image Processing XLII}, volume 11137,
  page 111370K. International Society for Optics and Photonics, 2019.

\bibitem{AlonsoSutterDeVergata}
T.~Alonso, G.~Sutter, and J.~E.~L. De~Vergara.
\newblock {LOCO-ANS}: An optimization of {JPEG-LS} using an efficient and
  low-complexity coder based on {ANS}.
\newblock {\em IEEE Access}, 9:106606--106626, 2021.

\bibitem{BlanesEtAL}
I.~Blanes, M.~Hernandez-Cabronero, J.~Serra-Sagrista, and M.~W. Marcellin.
\newblock Redundancy and optimization of {tANS} entropy encoders.
\newblock {\em IEEE Transactions on Multimedia}, pages 4341--4350, 2020.

\bibitem{Duda2021}
J.~Duda.
\newblock Encoding of probability distributions for asymmetric numeral systems.
\newblock {\em arXiv preprint arXiv:2106.06438}, pages 1--5, 2021.

\bibitem{MoffatPetri}
A.~Moffat and M.~Petri.
\newblock Large-alphabet semi-static entropy coding via asymmetric numeral
  systems.
\newblock {\em ACM Transactions on Information Systems}, 38(4):1--33, 2020.

\bibitem{WangWangLin}
N.~Wang, C.~Wang, and S-J. Lin.
\newblock A simplified variant of tabled asymmetric numeral systems with a
  smaller look-up table.
\newblock {\em Distributed and Parallel Databases}, 39(3):711--732, 2021.

\bibitem{YamamotoIwata}
H.~Yamamoto and K.~Iwata.
\newblock Encoding and decoding algorithms of {ANS} variants and evaluation of
  their average code lengths.
\newblock {\em arXiv preprint arXiv:2408.07322}, 2024.

\bibitem{Yokoo}
H.~Yokoo.
\newblock On the stationary distribution of asymmetric binary systems.
\newblock In {\em 2016 IEEE International Symposium on Information Theory
  (ISIT)}, pages 11--15. IEEE, 2016.

\bibitem{DubeYokoo}
H.~Yokoo and D.~Dub{\'e}.
\newblock Asymptotic optimality of asymmetric numeral systems.
\newblock In {\em Proc. 42nd International Symposium on Information Theory and
  Its Applications (SITA 2019)}, pages 26--29, 2019.

\bibitem{YamamotoIwata2}
H.~Yamamoto and K.~Iwata.
\newblock Asymmetric encoding-decoding schemes for lossless data compression.
\newblock 2026.

\bibitem{Bloom}
C.~Bloom.
\newblock {cbloom rants blog}.
\newblock \url{http://cbloomrants.blogspot.com}.
\newblock Accessed: 08.01.2022.

\bibitem{ColletBlog}
Y.~Collet.
\newblock {RealTime Data Compression blog}.
\newblock \url{http://fastcompression.blogspot.com}.
\newblock Accessed: 08.01.2022.

\bibitem{GiesenBlog}
F.~Giesen.
\newblock {The ryg blog}.
\newblock \url{https://fgiesen.wordpress.com}.
\newblock Accessed: 07.01.2022.

\bibitem{KosarajuManzini}
S.~Rao Kosaraju and Giovanni Manzini.
\newblock Compression of low entropy strings with {L}empel--{Z}iv algorithms.
\newblock {\em SIAM Journal on Computing}, 29(3):893--911, 1999.

\bibitem{Manzini}
Giovanni Manzini.
\newblock An analysis of the {B}urrows--{W}heeler transform.
\newblock {\em Journal of the ACM}, 48(3):407--430, 2001.

\bibitem{DubeYokooDist}
D.~Dub{\'e} and H.~Yokoo.
\newblock Fast construction of almost optimal symbol distributions for
  asymmetric numeral systems.
\newblock In {\em 2019 IEEE International Symposium on Information Theory
  (ISIT)}, pages 1682--1686. IEEE, 2019.

\bibitem{HowardVitter}
P.~G. Howard and J.~S. Vitter.
\newblock Analysis of arithmetic coding for data compression.
\newblock {\em Information processing \& management}, 28(6):749--763, 1992.

\bibitem{Strutz}
T.~Strutz.
\newblock Rescaling of symbol counts for adaptive {rANS} coding.
\newblock In {\em 2023 31st European Signal Processing Conference (EUSIPCO)},
  pages 585--589. IEEE, 2023.

\bibitem{Alverson}
R.~Alverson.
\newblock Integer division using reciprocals.
\newblock In {\em IEEE Symposium on Computer Arithmetic}, pages 186--190, 1991.

\bibitem{LemireBartlettKaser}
D.~Lemire, C.~Bartlett, and O.~Kaser.
\newblock Integer division by constants: optimal bounds.
\newblock {\em Heliyon}, 7(6), 2021.

\bibitem{CoverThomas}
T.~M. Cover and J.~A. Thomas.
\newblock Information theory and statistics.
\newblock {\em Elements of Information Theory}, 1(1):279--335, 1991.

\bibitem{Csiszar}
Imre Csisz{\'a}r.
\newblock The method of types.
\newblock {\em IEEE Transactions on Information Theory}, 44(6):2505--2523,
  2002.

\bibitem{Gagie}
Travis Gagie.
\newblock Large alphabets and incompressibility.
\newblock {\em Information Processing Letters}, 99(6):246--251, 2006.

\bibitem{McMillan}
B.~McMillan.
\newblock Two inequalities implied by unique decipherability.
\newblock {\em IRE Transactions on Information Theory}, 2(4):115--116, 1956.

\bibitem{AlonOrlitsky}
N.~Alon and A.~Orlitsky.
\newblock A lower bound on the expected length of one-to-one codes.
\newblock {\em IEEE Transactions on information theory}, 40(5):1670--1672,
  1994.

\bibitem{SzpankowskiVerdu}
W.~Szpankowski and S.~Verdu.
\newblock Minimum expected length of fixed-to-variable lossless compression
  without prefix constraints.
\newblock {\em IEEE Transactions on Information Theory}, 57(7):4017--4025,
  2011.

\bibitem{OrlitskySanthanam}
A.~Orlitsky and N.~P. Santhanam.
\newblock Speaking of infinity [iid strings].
\newblock {\em IEEE Transactions on Information Theory}, 50(10):2215--2230,
  2004.

\bibitem{Shtarkov}
Yu.~M. Shtar'kov.
\newblock Universal sequential coding of single messages.
\newblock {\em Problems of Information Transmission}, 23(3):3--17, 1987.

\bibitem{SzpankowskiWeinberger}
W.~Szpankowski and M.~J. Weinberger.
\newblock Minimax pointwise redundancy for memoryless models over large
  alphabets.
\newblock {\em IEEE Transactions on Information Theory}, 58(7):4094--4104,
  2012.

\bibitem{Townsend}
J.~Townsend.
\newblock A tutorial on the range variant of asymmetric numeral systems.
\newblock {\em arXiv preprint arXiv:2001.09186}, pages 1--11, 2020.

\bibitem{TurboRC}
{TurboRC: Turbo range coder + rANS}.
\newblock \url{https://github.com/powturbo/Turbo-Range-Coder}.
\newblock Accessed: 21.10.2023.

\bibitem{AlakuijalaBrotli}
J.~Alakuijala, A.~Farruggia, P.~Ferragina, E.~Kliuchnikov, R.~Obryk,
  Z.~Szabadka, and L.~Vandevenne.
\newblock Brotli: A general-purpose data compressor.
\newblock {\em ACM Transactions on Information Systems (TOIS)}, 37(1):1--30,
  2018.

\bibitem{PainskyRossetFeder}
A.~Painsky, S.~Rosset, and M.~Feder.
\newblock Large alphabet source coding using independent component analysis.
\newblock {\em IEEE Transactions on Information Theory}, 63(10):6514--6529,
  2017.

\end{thebibliography}

%\begin{IEEEbiographynophoto}{Dmitry Kosolobov}
%	Ph.D., senior researcher in Ural Federal University, Ekaterinburg, Russia.
%\vfill

\end{document}